\pdfoutput=1 
\documentclass[namedreferences]{solarphysics}
\usepackage[hyperref,optionalrh,solaromanenum]{spr-sola-addons} 
\usepackage{graphicx}                    
\usepackage{color}                       
\usepackage{breakurl}                         

\begin{document}

\begin{article}

\begin{opening}

\title{Using SDO/HMI magnetograms as a source of the solar mean magnetic field data}

%
\author[addressref=aff1,corref, email={alex.s.kutsenko@gmail.com}]{\inits{A.S.}\fnm{A.S.}~\lnm{Kutsenko}}
\author[addressref={aff1,aff2},email={vabramenko@gmail.com}]{\inits{V.I.}\fnm{V.I.}~\lnm{Abramenko}}

\address[id=aff1]{Crimean Astrophysical Observatory, Russian Academy of Science, Nauchny, Bakhchisaray, 298409, Crimea}
\address[id=aff2]{Central (Pulkovo) Astronomical Observatory, Russian Academy of Science (GAO RAN), Pulkovskoye ch. 65,  Saint-Petersburg, 196140, Russia}

%
\runningauthor{A.S.Kutsenko, V.I.Abramenko}
\runningtitle{Using SDO/HMI as a source of SMMF data}

\begin{abstract}

The solar mean magnetic field (SMMF) provided by the Wilcox Solar Observatory (WSO) is compared with the SMMF acquired by the \textit{Helioseismic and Magnetic Imager} (HMI) onboard the \textit{Solar Dynamic Observatory} (SDO). We found that despite the different spectral lines and measurement techniques used in both instruments the Pearson correlation coefficient between these two datasets equals 0.86 while the conversion factor is very close to unity: B(HMI) = 0.99(2)B(WSO). We also discuss artifacts of the SDO/HMI magnetic field measurements, namely the 12 and 24-hour oscillations in SMMF and in sunspots magnetic fields that might be caused by orbital motions of the spacecraft. The artificial harmonics of SMMF reveal significant changes in amplitude and the nearly stable phase. The connection between the 24-hour harmonic amplitude of SMMF and the presence of sunspots is examined. We also found that opposite phase artificial 12 and/or 24-hour oscillations exist in sunspots of opposite polarities.

\end{abstract}

%

\end{opening}

%
 \section{Introduction}
	 \label{sec-Introduction} 

The solar mean magnetic field (SMMF) represents imbalance of the line-of-sight (LOS) magnetic flux integrated over the entire visible solar hemisphere \citep{Garcia1999, Haneychuk2003}. A strong correlation exists between the SMMF and the interplanetary magnetic field measured from Earth \citep{Severny1970, Scherrer1977a, Bremer1996} as well as between the SMMF and the solar wind speed \citep{Neugebauer2000}. In a number of papers Sheeley \emph{et al.} \citep{Sheeley1985, Sheeley1986a, Sheeley1986b} investigated the evolution of photospheric magnetic fields and the connection between the SMMF and magnetic transport parameters, including differential rotation and the diffusion coefficient. Fitting the model parameters to obtain the best agreement between the simulations and the observed SMMF allowed them to confirm the existence of the meridional flow \citep{Wang1989} and to establish the influence of the differential rotation on the SMMF temporal variations.

Systematic measurements of the SMMF are carried out at the Crimean astrophysical observatory (CrAO) since 1968 \citep{Severny1969, Kotov1983}. Several observatories performed SMMF measurements during past four decades as well: Mount Wilson Observatory \citep{Howard1974}, Sayan Observatory of the Solar-Terrestrial Physics Institute \citep{Grigor'ev1987}, the Birmingham Solar Oscillation Network of observatories \citep{Chaplin2003}. The spaceborne \textit{Michelson Doppler Interferometer} (MDI, \citealp{Scherrer1995}) and \textit{Global Oscillations at Low Frequencies} (GOLF, \citealp{Garcia1999}) instruments onboard the ESA/NASA \textit{Solar and Heliospheric Observatory} (SOHO) produced high-cadence data on SMMF. The SOHO/MDI magnetic field observations were terminated in 2011. Instead the improved \textit{Helioseismic and Magnetic Imager} (HMI, \citealp{Scherrer2012}) onboard the \textit{Solar Dynamic Observatory} (SDO, \citealp{Pesnell2012}) began to provide data on solar magnetic fields. Daily SMMF measurements provided by Wilcox Solar Observatory (WSO) since 1975 \citep{Scherrer1977b} have become a recognized benchmark for this type of data. A brief comparison between the WSO and SOHO/MDI SMMF measurements was undertaken by \cite{Boberg2002}. 

In the present paper, we compare the SMMF obtained by the SDO/HMI instrument with that provided by WSO to ensure consistency of datasets from these two sources. We also discuss some issues, in particular, the 24-hour artificial periodicity of SDO/HMI SMMF.


 \section{Observables and methods}
	 \label{sec-Observ} 
In this research, we utilize two types of magnetic fields measurements. The first one is the WSO SMMF measurements that are carried on since May 1975 \citep{Scherrer1977b}. A Babcock-type magnetograph \citep{Babcock1953} is used in combination with a 33 cm coelostat and Littrow spectrograph \citep{Scherrer1977b}. The LOS component of the magnetic field is measured "as the difference between the Zeeman splitting in the Fe \small{I} line at 5250 \AA\ and the zero offset signal in the magnetically insensitive Fe \small{I} line at 5124 \AA" \citep{Boberg2002}. An observation run lasts twenty minutes and is repeated several times per day. The SMMF is computed by integration of the large-scale field over the solar disk. A daily averaged SMMF magnitude is then derived by averaging over all observations taken on a given day. The typical standard deviation is about 5 $\mu$T.

Our second data set consists of the SDO/HMI measurements. The spacecraft is located in an inclined geosynchronous orbit at the altitude of 36000 km. It outlines a figure "eight" trajectory along the $102^{\circ}$ W meridian spanning 57 degree in latitude as seen from the ground \citep{Smirnova2013}.  The HMI instrument \citep{Schou2012} onboard SDO is a filtergraph with the full-disk coverage by two CCD cameras of 4096$\times$4096 pixels \citep{Scherrer2012, Liu2012}. The cameras produce 12 filtegrams used to derive LOS magnetograms: the profile of the photospheric Fe \small{I} 6173 \AA\ spectral absorption line \citep{Norton2006} is measured at six wavelength positions in two polarization states. The magnetic field sign and the magnitude are derived from Zeeman splitting of the line. The LOS magnetograms are provided with 45 second and 720 second cadence. The 24-hour periodicities in the HMI magnetic field products were revealed by \cite{Liu2012}, which probably have resulted from the 24-hour cyclicity in the orbital velocity of the spacecraft combined with uncertainties in instrument calibration. The effect seems to be enhanced for strong fields related to sunspots \citep{Liu2012, Smirnova2013}.

We also used the CrAO daily patrol sunspots' magnetic field measurements \footnote{available at http://solar.craocrimea.ru/eng/observations.htm} to determine unsaturated peak magnitudes of the sunspot magnetic fields. Magnetic field strength in a sunspot is derived from the Zeeman splitting in the Fe \small{I} 6302 \AA\ line. The observations are presented as a solar disk sketch with sunspot group contours, their magnetic polarity, and field strength.

We calculated the SMMF magnitude by integrating the signed magnetic flux from all pixels of a 720-second magnetogram within a circumference with one solar radius centered at the solar disk center. The solar radius and the solar disk center data were taken from the header of the corresponding fits file.

\begin{figure}    
  	\centerline{\includegraphics[width=1.0\textwidth,clip=]{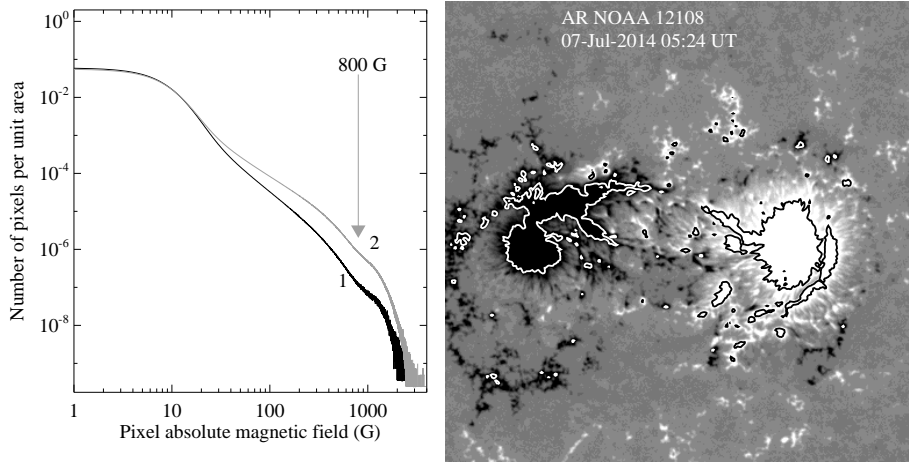}
  	}
  	\caption{Left - Typical distribution functions of HMI pixel values. Curve 1 was obtained as an average of 246 histograms from 720-second magnetograms acquired in 2010. Curve 2 is an average of 348 histograms from 720-second magnetograms acquired in 2014. One magnetogram per day was used in both cases. An arrow indicates the chosen 800 G weak-to-strong field threshold. Right - The HMI magnetogram of AR NOAA 12108. The thick contours mark the 800 G threshold.
  	}
  	\label{Fig1}
  \end{figure}

A typical histogram of the absolute magnetic field strength in HMI pixels is shown in Figure~\ref{Fig1}. The first curve (marked by 1 in the figure) was calculated as an average of 246 histograms calculated from full disk magnetograms measured between April 27 and December 31 of 2010. The second curve is the average of 348 histograms calculated from magnetograms acquired in 2014. One magnetogram per day was used in both cases. The histograms were normalized to unit area. The curves' behavior beyond 10 G is explained by approaching to the noise limit of the instrument. The first curve, associated with the period of an extended solar minimum, shows the lack of pixels of strong and medium fields compared to the second curve referring to the cycle maximum. Nevertheless, the slight bend in the histogram at 800 G is clearly seen. The right panel in Figure~\ref{Fig1} shows that the 800 G contour is predominantly a penumbra-to-umbra boundary. Accordingly we selected the 800 G value as a threshold to separate the sunspot's fields from the rest of the solar magnetic fields. Thus, we summed magnetic flux separately over the strong-field areas and over the rest of the solar disk. This allowed us to reveal the contribution of sunspots to the appearance of the 24-hour artificial periodicity. Throughout this paper these sums are referred to as weak- and strong-field components of the SMMF.

The Joint Science Operation Center (JSOC) \footnote{http://jsoc.stanford.edu/} run by the Stanford Solar Group allows to export LOS magnetogram FITS file headers alone that includes DATAMEAN parameter, which can be treated as SMMF. This parameter was used in this study to verify the calculated value of the total SMMF.

FITS file headers also contain the QUALITY parameter. A non-zero value of this parameter means that some data used in the magnetogram calculations were lost or corrupted. Such a magnetogram should not be used for scientific purposes. The HMI instrument calibration is performed at 06:00 UT and 18:00 UT every day and the corresponding magnetograms have non-zero QUALITY value. To minimize the influence of these bad data we used magnetograms obtained at twenty-fourth minute of every hour. Some of the selected magnetograms also had the non-zero QUALITY parameter, however in the cases the magnetogram showed no abnormal value of SMMF it was kept.

 
 \section{Results}
	 \label{sec-Results}
  \subsection{Comparing WSO SMMF with SDO/HMI SMMF}
	  \label{subs-Res1}
	  
\begin{figure}    
  	\centerline{\includegraphics[width=1.0\textwidth,clip=]{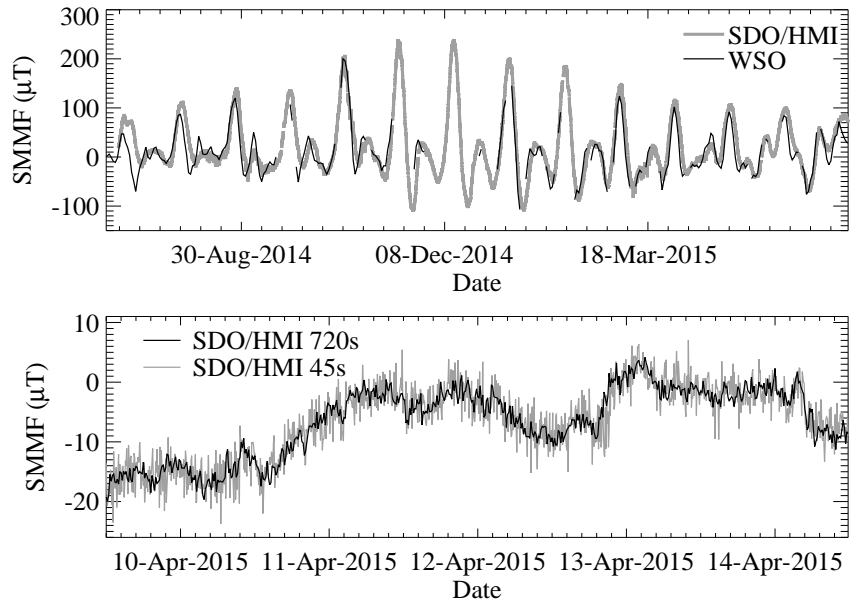}
  	}
  	\caption{Top - One-year WSO SMMF measurements (thin black line) overplotted with 720-s cadence SDO/HMI SMMF (thick grey line). WSO data points are obtained daily at 20:00 UT. Bottom - A five-day segment of the SDO/HMI SMMF derived from both 45-s and 720-s magnetograms. The standard deviation of the high-frequency noise of the 720-s magnetogram SMMF is 3 $\mu$T.}
  	\label{Fig2}
  \end{figure}

\begin{figure}    
  	\centerline{\includegraphics[width=0.5\textwidth,clip=]{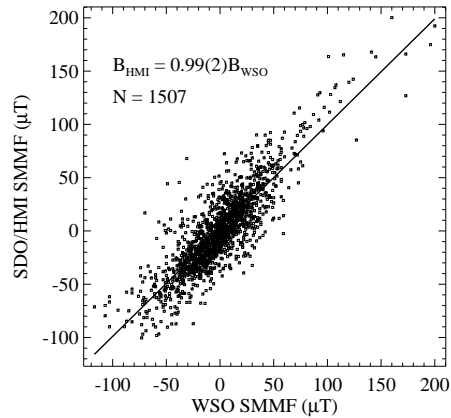}
  	}
  	\caption{Comparison of WSO SMMF and SDO/HMI SMMF measurements. The regression equation is B(HMI) = 0.99(2)B(WSO). Total 1507 data pairs were used. The Pearson correlation coefficient is 0.86.
  	}
  	\label{Fig3}
  \end{figure}

A one-year (July 2014 - June 2015) WSO and HMI SMMF data obtained by using the aforementioned approach is shown in Figure~\ref{Fig2}, top panel. Visual inspection revealed a very good correspondence between the measurements obtained with different instruments, which was further confirmed by calculations of the Pearson correlation coefficient, see Figure~\ref{Fig3}. The WSO SMMF data that were measured daily at 20:00 UT are compared with the co-temporal SDO/HMI SMMF data set. In total, 1507 data pairs were used to obtain the regression relationship between the WSO and SDO/HMI SMMF sets (Figure~\ref{Fig3}). Notwithstanding the fact that different photospheric lines as well as techniques were used for measurements, the linear regression coefficient is close to unity, namely 0.99(2), and the correlation coefficient is 0.86. Also, the advantage of a spaceborne instrument (weather, daylight, and seeing independence) can be seen in Figure~\ref{Fig2}: the gaps in WSO data can be successfully filled by uninterrupted HMI observations. A good example could be a rejuvenation of SMMF observed in December 2014. The effect was caused by a persistent emergence of magnetic flux in active regions, which significantly amplified the large-scale background fields \citep{Sheeley2015}. Unfortunately, the WSO data had many gaps in December 2014 right at the moments of the greatest increase of the absolute SMMF amplitude. Adopting the conversion factor as 1.0, we infer that the absolute maximum--minimum difference in SMMF was higher by approximately 40 \% than that derived from the WSO data only.

We estimated the level of high-frequency noise in the HMI SMMF time series. Low-frequency variations (13.5 and 27 day long and longer) were removed from the data by applying a high-pass filter to the SMMF data with a time constant of 24 hour. The standard deviation of the noise was found to be 3 $\mu$T for 720-second HMI magnetograms, which does not exceed the 5 $\mu$T value for the WSO data \citep{Scherrer1977b}. The noise level in both 45-second and 720-second HMI magnetograms can be visually estimated from the curves shown in the bottom panel of Figure~\ref{Fig2}: the ratio of SMMF noise level in 45-second magnetograms to that of 720-second magnetograms is close to 2.4. This value is in an agreement with the estimation of \cite{Liu2012} based on magnetograms acquisition time: the noise level ratio was reported to be about 2.48.

  \subsection{SDO/HMI SMMF artifacts}
	  \label{subs-Res2}

\begin{figure}    
  	\centerline{\includegraphics[width=1.0\textwidth,clip=]{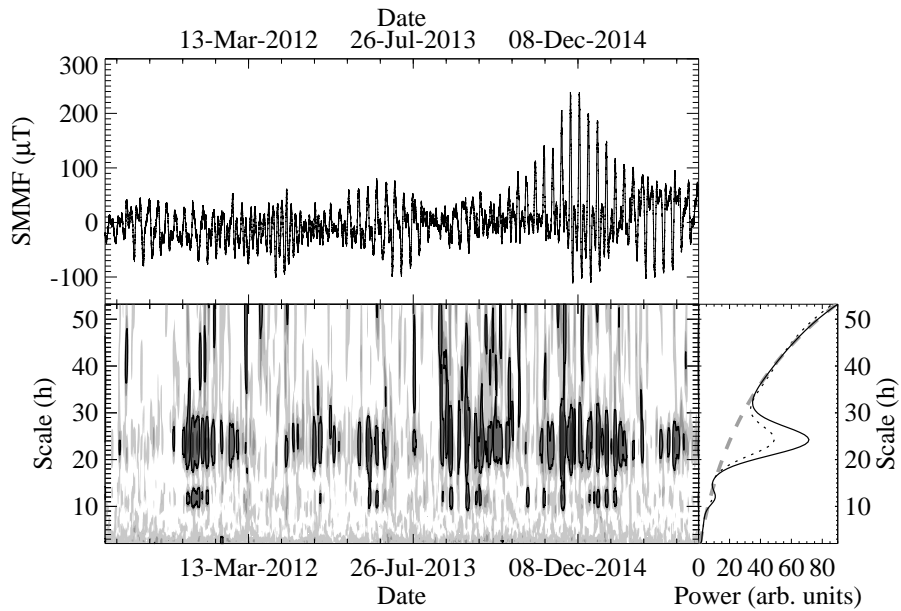}
  	}
  	\caption{Top - A five-year sequence of SDO/HMI SMMF measurements (January 01, 2011 - December 10, 2015). SMMF is calculated with a cadence of one hour. Bottom - the wavelet spectrum of the SMMF. The spectrum is averaged over the seven-day time interval. The scale covers the range from 2 to 52 hour. Thick black contours outline threefold excess of the harmonic power over the background level. The global wavelet spectrum is shown in the bottom right panel (solid black line). The 24-hour and 12-hour power peaks can be clearly seen. The empirical background level (dashed thick grey line, bottom right panel) was chosen as an interpolation of the global wavelet spectrum in the vicinity of 24-hour and 12-hour scales by an exponential fit. The global wavelet spectrum of the weak-field component is overplotted (dotted thin black line, bottom right panel).
  	}
  	\label{Fig4}
  \end{figure}

Using the technique described in Section 2, we calculated the SMMF components for a time interval from January 2011 to December 2015 with 1 hour cadence (Figure~\ref{Fig4}, top panel) to analyse artificial periodicities in SMMF.

The topic was discussed in a number of publications ({\it e.g.} \citealp{Liu2012, Smirnova2013}). The study by \cite{Liu2012} showed that the 12 and 24-hour magnetic field amplitude variations are present in SDO/HMI magnetograms and they can be attributed to the strong fields of active regions. The strong enough Zeeman splitting coupled with the Doppler effect due to the Sun rotation and the satellite movement caused the spectral line to shift every 24 hours beyond the well-determined part of the calibration curve. The calibration curve was extrapolated in this case and the periodicities are believed to be caused by some inaccuracy in the extrapolation. These 24-hour artifacts led to 3\% variations of the sunspot’s magnetic field amplitude \citep{Liu2012}. There is no evidence of significant variations of magnetic fields of quiet-Sun regions. \cite{Smirnova2013} found these artificial oscillations in both background and active region fields with the amplitude of the oscillations increasing rapidly when the field strength exceeded 2000 G (\cite{Smirnova2013}, see their conclusion).

To reveal the influence of the artifacts on SDO/HMI SMMF, we performed  a wavelet analysis \citep{Torrence1998} of the entire five-year long SMMF time series. A Morlet mother function with nondimensional frequency $\omega_0$ equal to six was chosen.

The wavelet transform of the SMMF is shown in the lower left panel of Figure~\ref{Fig4}. Both 24-hour and 12-hour peaks can be clearly seen on the global (averaged over the entire time interval) wavelet power plot (Figure~\ref{Fig4}, lower right panel). Since both 24-hour and 12-hour harmonics have an artificial origin, we interpolated the global wavelet spectrum in the vicinity of these harmonics by an exponential fit in order to obtain unperturbed SMMF spectrum (Figure~\ref{Fig4}, bottom right panel, grey thick dashed curve). We chose this curve as a spectrum of an empirical background. In the wavelet spectrum (Figure~\ref{Fig4}, bottom left panel) the solid thick curve outlines the threefold excess of the harmonic power over the background level. Analysis of the wavelet transforms of both weak- and strong-field components separately shows that only approximately 50\% of the spectral power of the 24-hour harmonic in the total SMMF is produced by the strong-field component (the global wavelet spectrum of the weak-field component is shown in the bottom right panel of Figure~\ref{Fig4}, black thin dotted curve). The Pearson correlation coefficient between the number of strong-field pixels and the amplitude of the 24-hour harmonic equals only 0.45.

\begin{figure}    
  	\centerline{\includegraphics[width=1.0\textwidth,clip=]{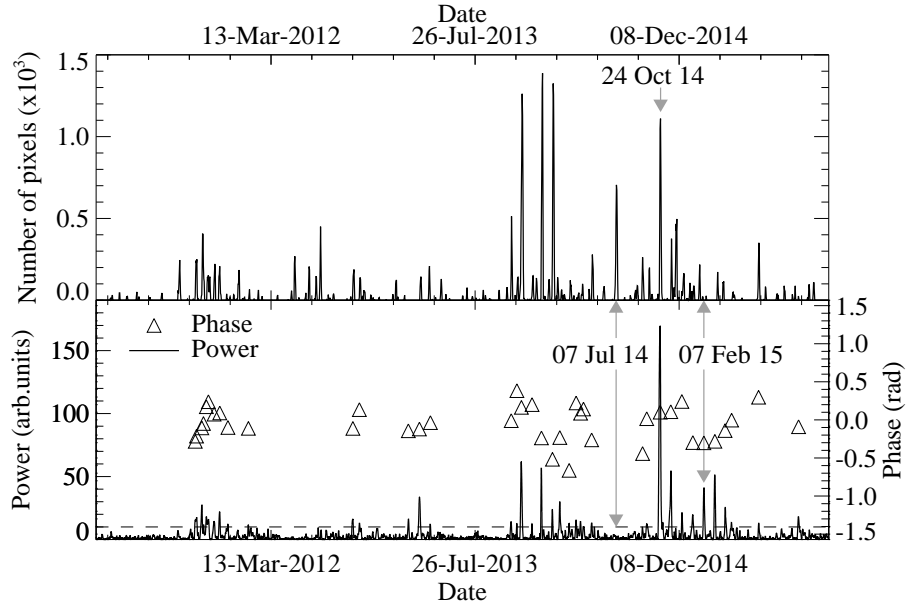}
  	}
  	\caption{Top - The variation of the number of pixels that have absolute magnetic field value exceeding 2000 G  during the five-year time interval. An emergence of NOAA AR12192 active region in October 2014 that caused a rapid notable increase of 24-hour harmonic amplitude is indicated by an arrow. Bottom - The amplitude of 24-hour harmonic. One can see that high number of strong-field pixels is not necessary associated with the increase of the 24-hour harmonic amplitude and vice versa (examples on July 14, 2014 and on February 07, 2015 are indicated by arrows). A moderate correlation exists between the curves (correlation coefficient is 0.55). The relative phase of the 24-hour harmonic is marked by triangles. The phase was calculated at the points where the harmonic amplitude exceeded the background level at least tenfold (dashed line).
  	}
  	\label{Fig5}
  \end{figure}

 The number of pixels that have the magnetic field strength above 2000 G is shown in Figure~\ref{Fig5}, top panel. The 2000 G level was suggested by \cite{Smirnova2013} as a threshold above which a rapid increase of the 24-hour harmonic amplitude was observed. In the same figure (bottom panel) the spectral power amplitude of 24-hour oscillations is shown for comparison. One can see that instances of a significant increase of the sunspot’s area are not necessarily  associated with an increase of the 24-hour oscillation amplitude (for example, a sunspot area peak on July 07, 2014). At the same time, a relatively high 24-hour harmonic amplitude can be observed in the absence of large sunspots ({\it e.g.}, harmonic peak on February 07, 2015). A moderate correlation exists between these two curves (Pearson’s correlation coefficient equals 0.55). This also confirms the conclusion that only approximately half of the 24-hour harmonic power is produced by strong-field solar regions. Meanwhile, both 24-hour and 12-hour harmonics were also detected in the global power spectrum of the weak-field component (Figure~\ref{Fig4}, bottom right panel, thin dotted black curve). In general, during the analysed five-year time period, the 24-hour harmonic amplitude exceeded the background level for 75\% of the time, whereas the triple background level was exceeded for 30\% of the time. Apparently, the persistence of the 24-hour harmonic in the weak-field component can be explained by the collective contribution of a very large number (more than 10 millions) of individual pixels that exhibited weak in-phase artificial variations of magnetic field that becomes significant after summation over the entire solar disk.
 
\begin{figure}    
  	\centerline{\includegraphics[width=1\textwidth,clip=]{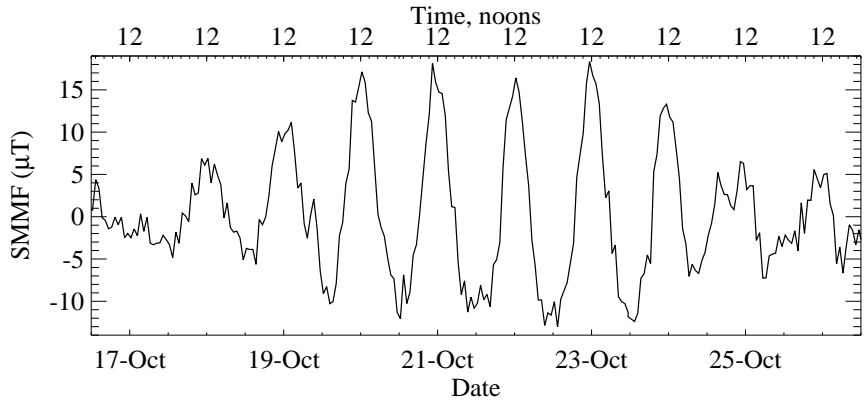}
  	}
  	\caption{A significant increase of 24-hour harmonic amplitude in October 2014. Only the high-frequency component of SMMF is shown. The peak-to-peak amplitude reaches values up to 30  $\mu$T. The perturbation of SMMF due to artificial 24-hour oscillation has extrema at 12:00 UT (positive) and at 00:00 UT (negative). The perturbation is negligible at 06:00 UT and at 18:00 UT when the relative orbital velocity of the SDO spacecraft is close to zero.
  	}
  	\label{Fig6}
  \end{figure}

Day-to-day variations of SMMF due to the 24-hour harmonic are very well pronounced in October, 2014. A rapid notable increase of the 24-hour harmonic amplitude from October 19 to October 25 (Figure~\ref{Fig5}) was caused by the emergence of an active region (AR) NOAA 12192 that culminated on October 23. It was the largest naked-eye sunspot group since November 1990 \citep{Sheeley2015}. The amplitude of the 24-hour oscillations during the aforementioned time period exceeded 20 $\mu$T  (Figure~\ref{Fig6}), which is nearly 10\% of the typical average level (200 $\mu$T) of the SMMF during a cycle maximum \citep{Garcia1999}.

The apparent regularity of the SMMF high-frequency variations in Figure~\ref{Fig6} motivated us to explore the phase of 24-hour harmonic (triangles in Figure~\ref{Fig5}, bottom panel). We derived the relative phase by the wavelet analysis at the moments when the 24-hour harmonic power exceeded the background level (bottom panel of Figure~\ref{Fig5}, dashed curve) at least tenfold. It was found that the phase remains quasi-stable during the five-year observation interval, which is expected assuming the artificial nature of the 24-hour harmonic. The phase deviation from the zero mean value does not exceed 0.5 rad and is caused by the data noise. The quasi-constant behavior of the phase explains why we observe artificial extrema in the SMMF at the same time instances around 12:00 UT (positive) and 00:00 UT (negative). The SMMF remains unperturbed at 06:00 UT and at 18:00 UT (except some specific days, see below). At these times the spacecraft velocity is close to zero \citep{Liu2016} that confirms once again the orbital origin of 24-hour harmonic.
	
\begin{figure}    
  	\centerline{\includegraphics[width=1\textwidth,clip=]{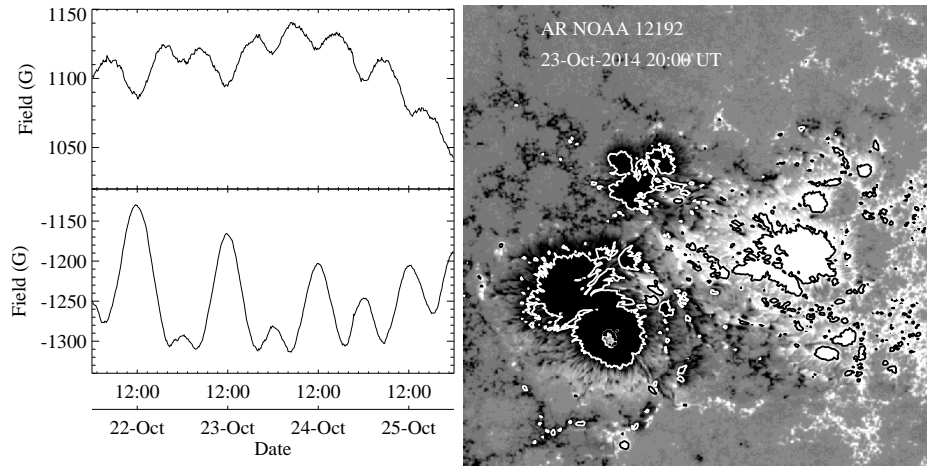}
  	}
  	\caption{Left - The magnetic fields of AR NOAA 12192 sunspots averaged over all the pixels that have value more than 800 G (upper panel) or less than -800 G (lower panel). The opposite phase oscillations of positive and negative fields are obvious. Right - A magnetogram of  AR NOAA 12192 on October 23, 2014. Thick contours 800 G outline areas over which the averaging was performed.
  	}
  	\label{Fig7}
  \end{figure}

	
\cite{Liu2012} pointed out that in addition to the uncertainties in the calibration curve, the pixel saturation can also be a source of artificial periodicities. Simulations showed that HMI measurements become saturated every 12 hours when the measured magnetic fields exceed 3200 G \citep{Liu2012}. AR NOAA 12192 data did exhibit the saturation in October 2014 \citep{Sun2015}. Thus, magnetic field strength in the umbra of the south-east core of the following spot reached values down to -3300 G, while the west leading spot umbra had approximately 2400 G (Figure~\ref{Fig7}) according to the CrAO data. We calculated the temporal variations of the average field of the spots during four days. Only pixels with the absolute magnetic field exceeding 800 G were summed, negative and positive field separately. Results are shown in Figure~\ref{Fig7}, left panel. Both 12 and 24-hour periodicities are clearly seen in the averaged magnetic field of the sunspots. Oscillations of positive and negative fields are in opposite phase that probably could be explained by saturation effect: the field cannot exceed the saturation level and it remains stable or diminishes.
	
\begin{figure}    
  	\centerline{\includegraphics[width=1\textwidth,clip=]{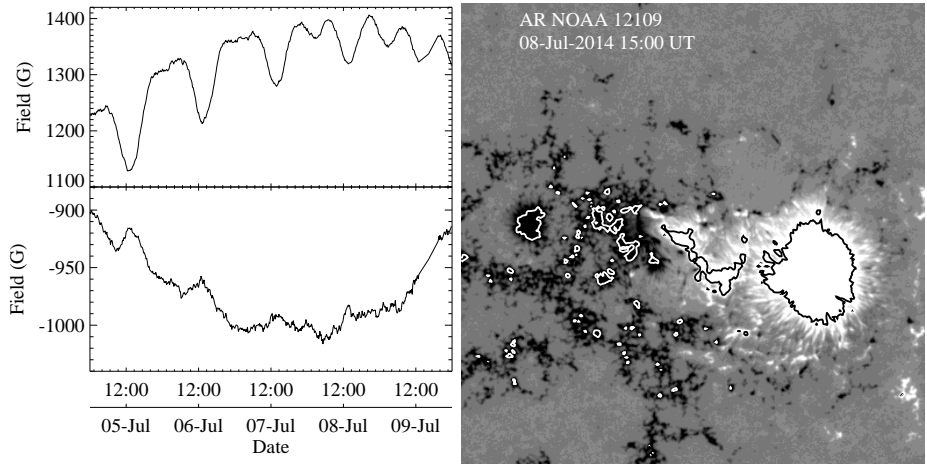}
  	}
  	\caption{The same as in Figure~\ref{Fig7} for AR NOAA 12109
  	}
  	\label{Fig8}
  \end{figure}

	
We performed the same procedure for a number of groups, namely AR NOAA 12032, 12033, 12034, 12108, 12109. The plot similar to Figure~\ref{Fig7} for AR NOAA 12109 is shown in Figure~\ref{Fig8}. The leading spot umbra of AR NOAA 12109 had magnetic field strength exceeding 3500 G while AR NOAA 12032, 12033, 12034 leading spots did not exceed -2500 G (CrAO), {\it i.e.} we analyzed both saturated and unsaturated sunspots. The results of the calculations are: a) artificial periodicity, if exists, can exhibit both 12 and 24-hour time period regardless of the strength of the spot; b) negative and positive sunspot fields oscillate in opposite phase; c) in general, stronger sunspots exhibit greater artificial harmonic amplitudes; d) amplitudes of 12/24-hour harmonics can vary spontaneously.

The first result makes us unable to distinguish the contribution of saturation from that of the calibration curve uncertainties. Both factors can act simultaneously and in the same manner. Since positive and negative pixels of AR data oscillate in opposite phase and the harmonic amplitudes can vary spontaneously, the total signed flux of ARs can reveal artificial oscillation of either of two harmonics (or their sum - 36-hour time period as detected by \cite{Smirnova2013}) or reveal no oscillations at all. The latter, probably, can explain the medium correlation between the sunspots area and the 24-hour harmonic amplitude and the possible absence of 24-hour harmonic in the SMMF in the presence of strong sunspots. On the other hand, the opposite phase oscillations of pixels of different polarity implies that the resulting phase of AR's magnetic field artificial periodicity can be determined by the dominant sunspot. This contradicts our above discussed result regarding the stable phase of the artificial periodicities in the SMMF. Changing the threshold value as well as integrating over constant size areas rather than over the pixels with certain values gave a similar outcome. Further investigations are needed to clarify the issue.

\begin{figure}    
  	\centerline{\includegraphics[width=0.5\textwidth,clip=]{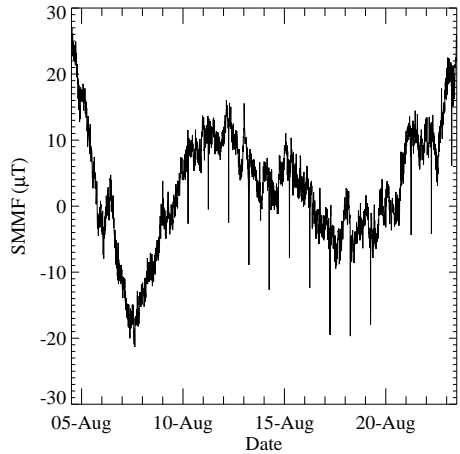}
  	}
  	\caption{A rapid artificial decrease of SMMF in August 2014 caused by calibration procedure. The decrease is observed from March 01 to 31 at 06:00 UT and from August 10 to September 12 at 18:00 UT. The mean magnitude of the decrease is 20 $\mu$T in March and 15 $\mu$T in August.
  	}
  	\label{Fig9}
  \end{figure}

We performed brief analysis of the non-zero QUALITY magnetograms acquired at the HMI calibration times at 6:00 and 18:00 UT. The calibration is performed every day at the moments when the orbital velocity is close to zero. Pixel-by-pixel comparison of three consequent magnetograms acquired at, say, 05:48 UT, 06:00 UT, and 06:12 UT, generally shows no noticeable significant discrepancies between the magnetic field values. However, for two times a year this is not true for statistical moments such as mean value. The peculiarity of the magnetograms derived at calibration times is a sudden decrease of the SMMF at certain days of year. Daily spikes are observed from March 01 to 31 at 6:00 UT and from August 10 to September 12 at 18:00 UT (Figure~\ref{Fig9}). The exception was the leap year 2012 when the spikes started to appear on February 29th. The mean absolute amplitude of the spikes is 20 $\mu$T in March and 15 $\mu$T in August. Due to the calibration process certain filtergrams are lost, which may cause a zero-point offset of weak field pixels \citep{Liu2016}. Summation over the entire solar disk (more than 10 million pixels) leads to significant error in the SMMF determination. The reason why the effect becomes significant only twice a year remains unclear.

 \section{Conclusions}
	  \label{sec-Conclusions}
		  
SDO/HMI instrument provides SMMF data with 45 s and 720 s cadence that are not subjected to natural seeing or daylight variations. The SMMF values are in a very good agreement with the WSO mean field data, and the Pearson correlation coefficient for the two data sets equals 0.86. Despite of the different techniques and spectral lines used for the measurements, the agreement between two datasets is impressive. The conversion factor is very close to unity (0.99(2)) and can be used to ensure the consistency between these two datasets.

Meanwhile, at least two types of artifacts exist in the SDO/HMI SMMF data. The first one is the 24-hour and 12-hour oscillations caused by the orbital motion of the spacecraft. The amplitude of the oscillations can vary significantly, can reach values up to tens of microtesla, and is related, in part, to the varying area of the sunspots on the disk. We found, however, that these artificial harmonics are presented in the SMMF power spectrum most of the time regardless of the presence of sunspots. All in all during the five-year time period of observations the 24-hour harmonic power exceeded the background level for 75\% of the time, whereas the triple background level was exceeded for 30\% of the time. The effect might be caused by the collective contribution of a large number (more than 10 millions) of individual pixels summed over the entire disk, each of them exhibiting in-phase weak artificial oscillations in the magnetic field amplitude. This is probably why the 24-hour and 12-hour oscillations did not get attention in previous studies \citep{Liu2012, Smirnova2013}, where the integration was performed over relatively small areas of the solar disk.

The phase of the 24-hour harmonic remains quasi-stable as expected from the artificial nature of 24-hour oscillations. The maximum positive and negative perturbations of the SMMF are observed at 12:00 UT and at 00:00 UT correspondingly.

%

%

%

%
 \begin{acks}
 	
 	
We thank Dr. V. Yurchyshyn for a careful reading of the manuscript and valuable suggestions and criticism that led to significant improvement of the manuscript. We also are grateful to Dr. Liu Yang for valuable information on SDO/HMI instrument, to V.I. Haneychuk for recommendation on SDO/HMI data processing. We thank the anonymous referee for his/her very useful comments and raised questions that helped us to improve the paper. SDO is a mission for NASA's Living With a Star (LWS) program. The SDO/HMI data were provided by the Joint Science Operation Center (JSOC). Wilcox Solar Observatory data used in this study was obtained via the website http://wso.stanford.edu, courtesy of J.T. Hoeksema. The wavelet software provided by C. Torrence and G. Compo (available at http://paos.colorado.edu/research/wavelets/) was useful for developing software applied in this paper. 

The reported study was supported by the RFBR research project 16-02-00221 A and the Presidium of the Russian Academy of Science Program 21.
 \end{acks}

%
%
%

\end{article} 
\end{document}